\documentclass[prb,twocolumn,a4paper,showpacs,aps]{revtex4-1}
\pdfoutput=1
\usepackage{amsmath}
\usepackage{amsfonts}
\usepackage{dsfont}
\usepackage{graphicx}
\usepackage{psfrag}
\usepackage{color}

\begin{document}

\title{Exchange cotunneling through quantum dots with spin-orbit coupling}
\author{J. Paaske, A. Andersen, and K. Flensberg}
\affiliation{The Niels Bohr Institute \& Nano-Science Center,
University of Copenhagen, Universitetsparken 5, DK 2100 Copenhagen,
Denmark}
\date{\today}

\begin{abstract}
We investigate the effects of spin-orbit interaction (SOI) on the
exchange cotunneling through a spinful Coulomb blockaded quantum
dot. In the case of zero magnetic field, Kondo effect is shown to
take place via a Kramers doublet and the SOI will merely affect the
Kondo temperature. In contrast, we find that the breaking of
time-reversal symmetry in a finite field has a marked influence on
the effective Anderson, and Kondo models for a single level. The
nonlinear conductance can now be asymmetric in bias voltage and may
depend strongly on direction of the magnetic field. A measurement of
the angle dependence of finite-field cotunneling spectroscopy thus
provides valuable information about orbital, and spin degrees of
freedom and their mutual coupling.
\end{abstract}

\maketitle


Quantum dots based on materials with pronounced spin-orbit
interaction (SOI), such as InAs, SiGe, carbon nanotubes, and single
molecules have recently received reinforced
attention.~\cite{Fasth07, Jespersen06, Csonka08, Takahashi09,
Katsaros10, Kuemmeth08, Galpin10, Fang08, Herzog10} This is
partially motivated by the quest for achieving electrical control of
single spins, utilizing the fact that an electrical coupling to the
{\it orbital} degrees of freedom may allow for manipulations of the
{\it spin} via the
SOI.~\cite{Levitov03,Stepanenko04,Debald05,Flindt06,Nowack07,Trif07}
In quantum dots, the precise form of the spin-orbit coupling depends
strongly on the band structure, confining potential and dot geometry
altogether. It would therefore be of great value if one could infer
about the SOI directly from a measured cotunneling
bias spectroscopy, which is known to produce sharp spectroscopic
features due to threshold processes and/or Kondo effects.

It is well known that Kondo effect in metals with magnetic
impurities like Ce and Yb, say, are strongly affected by spin-orbit
interaction~\cite{Coqblin69,Nozieres80,Yamada84}. The SOI modifies
the spectrum of the impurity atom, but since it preserves
time-reversal invariance a Kramers-degenerate groundstate remains
and gives rise to Kondo effect. Likewise, a quantum dot holding a
net spin-1/2, will also have its spectrum modified by SOI, and a
Kramers degeneracy will still be available for Kondo effect. Unlike
the atomic $\mathbf{L}\cdot\mathbf{S}$ coupling, however, the SOI in
a quantum dot breaks rotational invariance and relates to specific
spatial directions, akin to the effect of a crystal
fields~\cite{Coqblin69,Nozieres80,Yamada84} or nearby
surfaces~\cite{Ujsaghy98,Szunyogh06} in the atomic problem. Since a
spinful quantum dot allows for local directional probes such as bias
voltage and magnetic field, the question arises if there are effects
of SOI that show up directly in a transport measurement?


Here we show that in the case where a single level approximation is
valid, the SOI can be absorbed in a redefinition of the lead electron
fields and thus leaves the Kondo effect unaffected. In the presence
of a finite magnetic field, however, spin, and orbital contents of
the Kramers doublets become disentangled and a spatial asymmetry in
the tunneling amplitudes can cause the Zeeman-split Kondo peak to
become asymmetric in bias voltage. This type of asymmetric splitting
does not occur without SOI, unless the voltage becomes large enough
to allow for real charge fluctuations on the dot. Furthermore, the
SOI induced asymmetry will depend strongly on the direction of the
magnetic field. The distinct angular dependence provides a very
direct signature of the SOI in a quantum dot, thus providing
valuable information about the quantum dot in question.


We employ the following general single-particle Hamiltonian to
describe a quantum dot defined by a potential $V(\mathbf{r})$ and
placed in an external magnetic field
\begin{align}
\mathcal{H}_{d}=
&\frac{({\mathbf{p}}\!-\!e{\mathbf{A}})^{2}}{2m}+V({\mathbf{r}})
+g\mu_{B}{\mathbf{B}}\cdot{\boldsymbol{\tau}}  \notag \\
&\hspace*{10mm}+ \frac{e\hbar}{4m^{2}c^{2}}
[\mathbf{{E}(r)\times(\mathbf{p}-e%
\mathbf{A})]\cdot{\boldsymbol{\tau}},}\label{eq:SPH}
\end{align}
with ${\boldsymbol{\tau}}$ denoting the vector of Pauli matrices,
and $\mathbf{B}$ the external magnetic field corresponding to a
vector potential $\mathbf{A}$. The spin-orbit coupling is here kept
on its most generic form in terms of the relevant nuclear or
structural electrical field $\mathbf{E}({\mathbf{r}})$. The
potential contains both the periodic potential from the ionic
background and the imposed confining potentials defining the dot.


In the absence of an external field $({\mathbf{A}}={\mathbf{0}})$,
this Hamiltonian is symmetric under time reversal and its
eigenstates therefore take the form of Kramers doublets of
two spinors~\cite{Sakurai94}:
\begin{equation}
\psi _{n\Uparrow }(\mathbf{{r})=\left(
\begin{array}{c}
u_{n}({r}) \\
v_{n}({r})%
\end{array}%
\right) ,\,\,\,\,\psi _{n\Downarrow }({r})=\left(
\begin{array}{c}
-v_{n}^{\ast }({r}) \\
u_{n}^{\ast }({r})
\end{array}
\right) ,}\label{eq:kramerspair}
\end{equation}
where the wavefunction components $u_{n}$ and $v_{n}$ depend
strongly on the confining potential. The corresponding
eigenenergies, $\varepsilon_{n}$, come with a characteristic
level spacing set by the confining potential and the strength of the
SOI. Also the source, and drain electrodes may experience a SOI, so
in general, we can express the eigenstates of the corresponding
Hamiltonian in the leads, $\mathcal{H}_{L/R}$, in the same way:
\begin{equation}
\psi _{\alpha \mathbf{k}\uparrow }(\mathbf{{r})=}\left(
\begin{array}{c}
a_{\alpha \mathbf{k}}(\mathbf{r}) \\
b_{\alpha \mathbf{k}}(\mathbf{r})
\end{array}
\right)\mathbf{,\,\,\,\,}
\psi\mathbf{_{\alpha \mathbf{k}\downarrow }({r})=}
\left(
\begin{array}{c}
-b_{\alpha \mathbf{k}}^{\ast }(\mathbf{r}) \\
a_{\alpha \mathbf{k}}^{\ast }(\mathbf{r})
\end{array}
\right),
\end{equation}
where $\alpha =L,R$ refers to left, and right lead, respectively.
Using these eigenstates, the total many-body Hamiltonian is given by
\begin{align}
H=&\sum_{\overset{\alpha =L/R}{\mathbf{k},\,\nu }}
(\varepsilon_{k}-\mu_{\alpha })
c_{\alpha \mathbf{k}\nu }^{\dagger }c_{\alpha\mathbf{k}\nu}
+\sum_{n,\eta }\varepsilon_{n}d_{n\eta }^{\dagger }d_{n\eta }\\
&
+\sum_{\overset{\alpha =L/R}{\mathbf{k},\,\nu,\,\eta,n}}
\left(
t^{\nu\eta}_{\alpha kn}
c_{\alpha\mathbf{k}\nu }^{\dagger }d_{n\eta }
+t^{\eta\nu}_{n\alpha k}
d_{n\eta}^{\dagger}c_{\alpha\mathbf{k}\nu}\right)+H_{int},
\notag\label{eq:Ham}
\end{align}
where $c_{\alpha k\nu }^{\dagger }$ creates an electron in the
$\nu$'th component of the Kramers doublet
$(\nu=\uparrow/\downarrow)$ with momentum $\mathbf{k}$ in lead
$\alpha=L/R$, and $d_{n\eta}^{\dagger }$ creates an electron in the
$\eta$'th component $(\eta=\Uparrow/\Downarrow )$ of the $n$'th
Kramers doublet on the dot. For the interaction term we employ the
constant interaction model $H_{int}=E_{C}(N-N_{g})^{2}$, where
$E_{C}$ denotes the total capacitive charging energy of the dot.

The amplitudes for tunneling between dot and leads $t^{\alpha k
n}_{\nu\eta}$ depend on the index of the Kramers doublets and it is
given by the Hamiltonian overlap,
\begin{equation}
t^{\alpha\mathbf{k}n}_{\nu\eta}=\int\!\!d{\mathbf{r}}\,\,
\psi^{\ast}_{\alpha\mathbf{k}\nu}(\mathbf{r})
\mathcal{H}_{tot}(\mathbf{r})
\psi_{n\eta}(\mathbf{r}),
\end{equation}
where the total single-particle Hamiltonian still takes the form of
Eq.~\eqref{eq:SPH}, but with an extended potential defining two
tunneling barriers which support the distinction into leads and dot
($\mathcal{H}_{tot}=\mathcal{H}_{d}+\mathcal{H}_{L}+\mathcal{H}_{R}$)
made in our definition of the eigenfunctions for the separate parts.
Regardless of the details of this potential, this first-quantized
Hamiltonian takes the following form:
\begin{equation}
\mathcal{H}_{tot}(\mathbf{r})=\mathcal{H}_{0}(\mathbf{r})\tau^{0}+
i\lambda_{so}\varepsilon_{ijk}\tau^{i}E_{j}(\mathbf{r})
\partial_{x_{k}},
\end{equation}
with kinetic energy and local potential contained in
$\mathcal{H}_{0}(\mathbf{r})$ and the \textit{local} spin-orbit term
written in terms of the Levi-Cevita symbol $\varepsilon_{ijk}$
(Einstein summation convention implied). Using the fact that the
different Kramers doublet components can be related via
time reversal~\cite{Sakurai94}, i.e.
\begin{equation}
\psi_{n\Uparrow,\sigma}=
i\tau^{y}_{\sigma\sigma^{\prime}}
\psi^{\ast}_{n\Downarrow,\sigma^{\prime}},
\end{equation}
together with the relation
$\tau^{y}\tau^{i}\tau^{y}=-(\tau^{i})^{\ast}$, it is readily
demonstrated that
\begin{align}
\langle\alpha\mathbf{k}\downarrow|\mathcal{H}_{tot}|
n\Downarrow\rangle=&\,\langle \alpha\mathbf{k}\uparrow|\mathcal{H}_{tot}|
n\Uparrow\rangle^{\ast},\\
\langle\alpha\mathbf{k}\downarrow|\mathcal{H}_{tot}|
n\Uparrow\rangle=&-\langle \alpha\mathbf{k}\uparrow|\mathcal{H}_{tot}|
n\Downarrow\rangle^{\ast},
\end{align}
which renders the tunneling amplitude proportional to a unitary
matrix in $\nu\eta$ space:
\begin{equation}\label{eq:tU}
t^{\alpha\mathbf{k}n}_{\nu\eta}=
t_{\alpha\mathbf{k}n}\mathbb{U}^{\alpha{\mathbf{k}n}}_{\nu\eta}.
\end{equation}

Next, we consider a specific charge state with an odd number of
electrons on the dot and assume all levels below the $m$'th level to
be doubly occupied. For the singly occupied $m$'th level the
dimensionless unitary matrix in Eq.~\eqref{eq:tU} can be now be
absorbed in a redefinition of the fermion fields in the two leads:
${\tilde c}^{\dagger}_{\alpha\mathbf{k}\eta}=
c^{\dagger}_{\alpha\mathbf{k}\nu}
\mathbb{U}^{\alpha{\mathbf{k}m}}_{\nu\eta}$. For sufficiently large
level spacings, we thus end up with the following single-orbital
Anderson model:
\begin{align}
H=&\sum_{\alpha\mathbf{k}\eta}(\varepsilon_{k}-\mu_{\alpha})
{\tilde c}^{\dagger}_{\alpha\mathbf{k}\eta}
{\tilde c}_{\alpha\mathbf{k}\eta}+
\sum_{\eta}\varepsilon_{m}d_{m\eta}^{\dagger}d_{m\eta}\notag\\
&+\sum_{\alpha\mathbf{k}\eta}t_{\alpha \mathbf{k}m}
\left({\tilde c}_{\alpha\mathbf{k}\eta}^{\dagger}d_{m\eta}+
d_{m\eta}^{\dagger}{\tilde c}_{\alpha\mathbf{k}\eta}\right)+
H_{int},\label{eq:Hamfin}
\end{align}
which no longer bears any trace of the SOI. Notice that this unitary
transformation is specific to the $m$'th level and therefore
tunneling amplitudes to any of the other levels on the dot will in
general retain their full (unitary) $2\times 2$ matrix structure in
$\nu\eta$ space. Apart from its influence on the precise magnitude
of $t_{\alpha\mathbf{k}m}$, SOI thus appears to have no effect
whatsoever on transport phenomena involving only a single level. In
particular, the Kramers degeneracy of this level will give rise to
Kondo effect.


This conclusion changes dramatically in the case of a finite applied
magnetic field, which couples directly to the constituent quantum
numbers of the Kramers doublets, i.e. to {\it spin}, and {\it
orbital} degrees of freedom. Using symmetric gauge, ${\bf A}({\bf
r})=({\bf B}\times{\bf r})/2$, the magnetic field enters
$\mathcal{H}_{d}$ through the kinematic momentum. This gives rise to
the following first quantized terms:
\begin{align}
\mathcal{H}_{B}=&-\mu_{B}{\bf B}\cdot{\bf L}
+\mu_{B}\frac{e}{4}(r^{2}B^{2}-({\bf r}\cdot{\bf B})^{2})
\label{eq:HB}\\
&+\mu_{B}\left[g{\bf B}
+\left(({\bf E}({\bf r})\cdot{\bf B}){\bf r}
-({\bf E}({\bf r})\cdot{\bf r}){\bf B} \right)
\frac{e}{4mc^{2}}\right]\cdot{\boldsymbol\tau},\nonumber
\end{align}
with an orbital term depending on the angular momentum operator
${\bf L}$, a diamagnetic term quadratic in $B$, and a local
anisotropic Zeeman term. The terms linear in $B$ both break the
time-reversal symmetry and thus destroy the degeneracy of the
Kramers doublets. We shall assume $B$ to be weak enough that this
splitting, which we parameterize by an effective $g$ factor,
${\tilde g}$, is much smaller than the relevant zero-field
level spacing, i.e. \hbox{${\tilde
g}\mu_{B}B\ll\Delta\varepsilon\equiv
\min[\varepsilon_{m+1}-\varepsilon_{m},\varepsilon_{m}-
\varepsilon_{m-1}]$}.

Apart from this renormalization of the Zeeman splitting within the
$m$'th level, the linear terms in $B$ also have off-diagonal terms
which couple the state $|m\eta\rangle$ to other states
$|n\eta'\rangle$ via ${\bf L}$ and ${\boldsymbol\tau}$. The
amplitudes for tunneling into the resulting finite $B$ eigenstates
of the dot are therefore changed and in particular the unitarity of
$t_{\alpha{\bf k}m}^{\eta'\!\eta}$ used for $B=0$ is no longer
guaranteed. In general, the matrix of tunneling amplitudes can be
polar decomposed into a product of a unitary and a hermitian matrix.
The unitary part can again be absorbed in a canonical transformation
of the conduction electrons in the corresponding lead and $t_{\alpha
{\bf k}m}^{\eta'\!\eta}$ can be taken to be hermitian in
$\eta'\!\eta$ space. Altogether, the tunneling term in
(\ref{eq:Hamfin}) is therefore modified to
\begin{align}
H_{T}=&\sum_{\alpha{\bf k}\eta'\!\eta} \left(t_{\alpha {\bf
k}m}^{\eta'\!\eta} {\tilde c}_{\alpha{\bf
k}\eta'}^{\dagger}\tilde{d}_{m\eta}+ (t_{\alpha {\bf
k}m}^{\eta'\eta})^{\!\ast} \tilde{d}_{m\eta}^{\dagger}{\tilde
c}_{\alpha{\bf k}\eta'}\right),
\end{align}
where electron creation operators, $\tilde{d}_{m\eta}^{\dagger}$ and
${\tilde c}_{\alpha{\bf k}\eta'}^{\dagger}$, as well as tunneling
amplitudes $t_{\alpha {\bf k}m}^{\eta'\!\eta}$ now depend on the
applied magnetic field.

{\it Kondo model:} Within the Kondo regime,
$\max(\nu_{F,\alpha}|(t_{\alpha'm}^{\eta\eta'})^{\ast} t_{\alpha
m}^{\eta''\eta'''}|)\ll\min(-\varepsilon_{m},\varepsilon_{m}+E_{C})$,
a Schrieffer-Wolff transformation~\cite{Schrieffer66} with the full
$\eta'\eta$ matrix tunneling amplitudes now leads to the following
exchange-cotunneling (Kondo) model:
\begin{align}
H_{K}=&\sum_{\alpha{\bf k}\eta} (\varepsilon_{k}-\mu_{\alpha})
{\tilde c}^{\dagger}_{\alpha{\bf k}\eta} {\tilde c}_{\alpha{\bf
k}\eta}+\mu_{B}{\tilde g}^{ij}B^{i}S^{j}
\label{eq:HamCotun}\\
&\hspace{10mm}+\frac{1}{2}\sum_{\stackrel{\alpha'\alpha,{\bf k'}{\bf
k},\eta'\eta} {i,j=0,x,y,z}} J^{ij}_{\alpha'\alpha}S^{i}\,{\tilde
c}^{\dagger}_{\alpha'{\bf k}'\eta'} \tau^{j}_{\eta'\eta} {\tilde
c}_{\alpha{\bf k}\eta},\nonumber
\end{align}
with $S^{0}=\mathds{1}$, $\tau^{0}_{\eta'\eta}=\delta_{\eta'\eta}$
and cotunneling amplitudes:
\begin{equation}
J^{ij}_{\alpha'\alpha}={\rm Tr}[t_{\alpha'm}\tau^{i}t_{\alpha
m}^{\dagger}\tau^{j}]
\frac{\varepsilon_{+}+(-1)^{\delta_{i0}}\varepsilon_{-}}
{2^{(\delta_{i0}+\delta_{j0})}\varepsilon_{+}\varepsilon_{-}},
\label{eq:amp}
\end{equation}
where $\varepsilon_{\pm}$ denotes the addition, and subtraction
energies on the dot.
Away from the particle-hole symmetric point,
$\varepsilon_{+}=\varepsilon_{-}$, a {\it vector} of
potential scattering amplitudes, $J^{0j}$, is present. Notice also
that expanding (\ref{eq:amp}) to leading order in $B$, it follows
from the hermiticity of $t_{\alpha{\bf k}m}^{\eta'\!\eta}$ that the
{\it intra}-lead exchange couplings will be diagonal and isotropic,
i.e. $J^{ij}_{\alpha'\alpha}\propto\delta_{ij}$.

It is interesting to note that this exchange scattering has lead
indices ($L/R$) mixed up with Kramers doublet indices ($\eta$) in
such a way that the usual simplification to a single channel Kondo
model no longer is possible. For $B=0$ (or without SOI) only one
channel is involved, but a finite field breaks the L/R symmetry via
the SOI and gives rise to a channel asymmetric two-channel
(anisotropic) Kondo model. As we shall demonstrate below, certain
system geometries will have {\it zero} Zeeman splitting in {\it
finite} magnetic field and therefore a strong coupling two-channel
regime~\cite{Rosch01,Pustilnik04} should in fact be attainable for
such geometries.

{\it Cotunneling current:} From these cotunneling amplitudes one can
now calculate the current through the dot as~\cite{Bruus04}
$I=e\sum_{\eta'\eta}(\Gamma^{RL}_{\eta'\eta}
-\Gamma^{LR}_{\eta'\eta})P(\eta)$, The non-equilibrium occupation
numbers $P(\eta)$ for the dot states satisfy a rate equation from
which they are found to be
$P(\eta)=\Gamma_{\eta\bar{\eta}}/(\Gamma_{\eta\bar{\eta}}
+\Gamma_{\bar{\eta}\eta})$, with
$\Gamma_{\eta'\eta}=\sum_{\alpha'\alpha}
\Gamma^{\alpha'\alpha}_{\eta'\eta}$. These cotunneling rates are
found as $\Gamma^{\alpha'\alpha}_{\eta'\eta}=
\gamma^{\alpha'\alpha}_{\eta'\eta}
\Delta\varepsilon^{\alpha'\alpha}_{\eta'\eta}
n_{B}(\Delta\varepsilon^{\alpha'\alpha}_{\eta'\eta})$ where $n_{B}$
is the Bose function and the energy differences are defined as
$\Delta\varepsilon^{\alpha'\alpha}_{\eta'\eta}=
\tilde{\varepsilon}_{m,\eta'}
-\tilde{\varepsilon}_{m,\eta}-\mu_{\alpha}+\mu_{\alpha'}$. Finally,
the tunneling probabilities are $\gamma^{\alpha'\alpha}_{\eta'\eta}=
\pi\nu_{F,\alpha'}^{{}}\nu_{F,\alpha}^{{}} \sum_{ijk}
J^{ij}_{\alpha'\alpha}J^{ik}_{\alpha\alpha'}
\tau^{j}_{\eta'\eta}\tau^{k}_{\eta\eta'}$. Notice that this is a
real number since
$(J^{ij}_{\alpha'\alpha})^{\ast}=J^{ij}_{\alpha\alpha'}$. As for a
system without SOI, the nonlinear conductance will exhibit cusped
steps at bias voltage, $V=\mu_{L}-\mu_{R}$, corresponding to the
effective Zeeman splitting. Since, however, Kramers doublet and lead
indices are mixed for finite magnetic field, the nonlinear
conductance is no longer symmetric in bias voltage. In general, the
two cusps at respectively positive and negative bias can now be of
different magnitude, and their relative magnitude will in general
depend on the angle of the magnetic field.

\begin{figure}[tb]
\centerline{\includegraphics[width=\columnwidth]{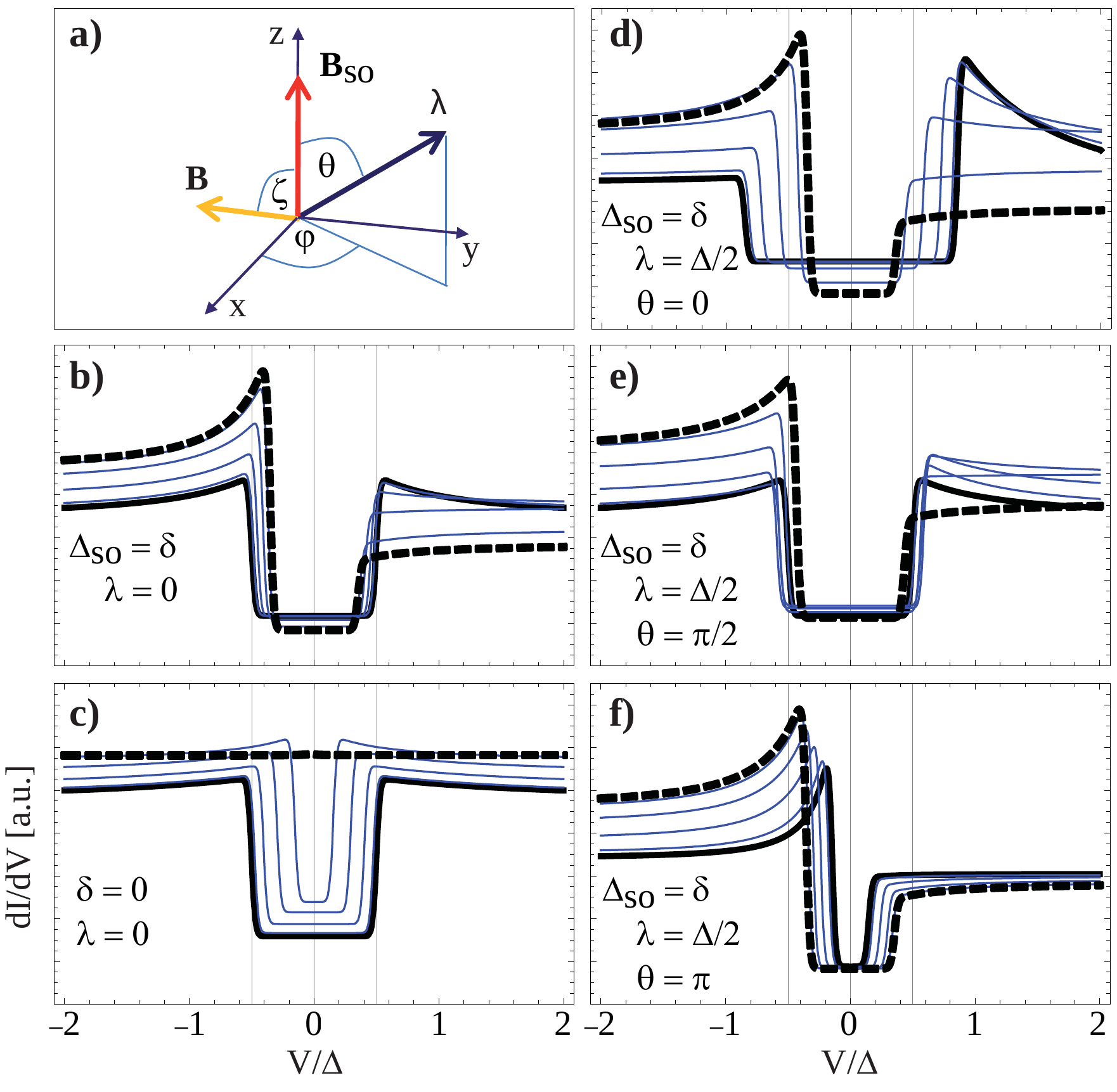}}
\caption{(a) Geometry of vectors $\boldsymbol{\protect\lambda}$,
$\mathbf{B}_{SO}$, and $\mathbf{B}$, indicating their relative
angles. $\mathbf{B}_{SO}$ and $\mathbf{B}$ span the $x-z$ plane.
(b-f) Nonlinear conductance, $dI/dV$, in arbitrary units vs.
bias voltage in units of $\Delta$ for a set of representative
parameters given in the insets. Each panel shows a progression of
curves with varying $\zeta$, moving from the thick solid (black)
curve with $\zeta=0$ to the thick dashed (black) curve with
$\zeta=\pi/2$. Remaining parameters are $g\mu_{B}B=0.5\Delta$,
$\varphi=\pi/4$, $t_{1,L/R}=1$, $t_{2,L}=3$, $t_{2,R}=0.1$. Notice
the closing of the Zeeman splitting for
$\mathbf{B}\perp\mathbf{B}_{SO}$ for two solely SOI-split orbitals
in panel (c). This can also be seen analytically from
Eq.~\eqref{eq:Z}.} \label{fig:CondPlots}
\end{figure}

{\it Two-level model:} To better illustrate these results, we now
exemplify our discussion by a simple two-level model (cf.
Fig.~\ref{fig:ModelFig}). With two levels split by an energy
$\delta$, we can express the Hamiltonian in the basis $\{|1\uparrow
\rangle$, $|1\downarrow\rangle$, $|2\uparrow\rangle$,
$|2\downarrow\rangle\}$, where the wave functions of the two levels,
$\psi _{1}$ and $\psi_{2}$, are chosen to be real. In this basis,
the spin-orbit coupling is included to give the full
dot Hamiltonian:
\begin{figure}[t]
\centerline{\includegraphics[width=0.7\columnwidth]{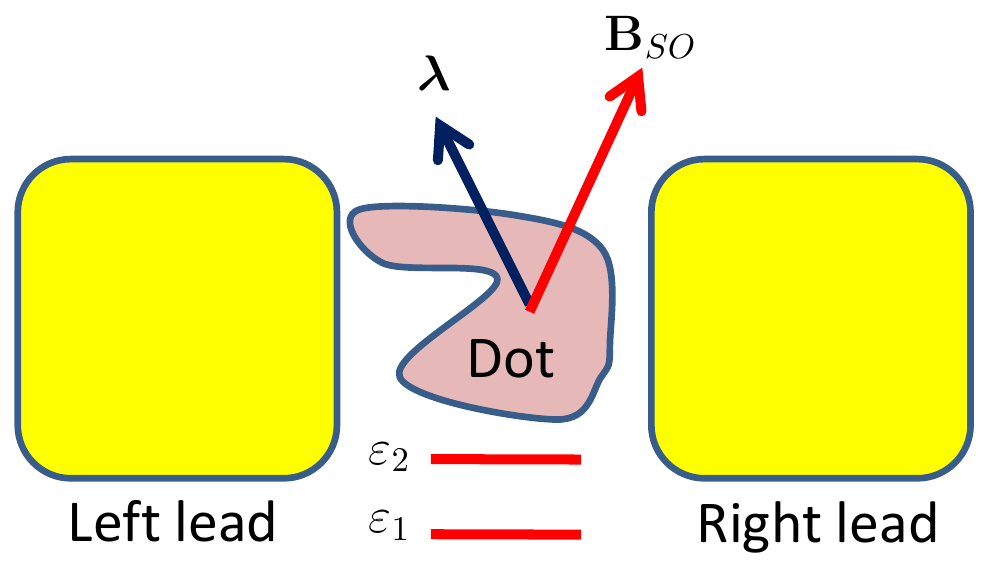}}
\caption{Sketch of two-orbital model system characterized by
spin-orbit field $\mathbf{B}_{SO}$ and the angular momentum vector
$\boldsymbol{\protect\lambda}$.} \label{fig:ModelFig}
\end{figure}
\begin{equation}
\langle i\sigma|H_{0}|j\sigma'\rangle= -\tau^{z}_{ij}
\tau^{0}_{\sigma\sigma'}\delta/2
+\tau^{y}_{ij}\tau^{z}_{\sigma\sigma'}\Delta_{SO}/2,
\end{equation}
where we have chosen the spin quantization along the built-in
spin-orbit field, $\mu_{B}\mathbf{B}_{SO}=\frac{e\hbar }{2mc^{2}}
\langle \psi _{1}|\mathbf{E}(\mathbf{r})\times\mathbf{p}|\psi_{2}
\rangle\equiv -\mathbf{\hat{z}~}i\Delta_{SO}/2$, characteristic for
these two levels. $H_{0}$ is diagonalized by two Kramers doublets:
\begin{subequations}
\label{eq:ab0}
\begin{eqnarray}
|a\eta\rangle&=&u|1\eta\rangle-iv\tau^{z}_{\eta\eta}|2\eta\rangle,\\
|b\eta\rangle&=&v|1\eta\rangle+iu\tau^{z}_{\eta\eta}|2\eta\rangle,
\end{eqnarray}
\end{subequations}
with $u^2+v^2=1$, $2uv=\Delta_{SO}/\Delta$, $\Delta=\sqrt{\delta
^{2}+\Delta _{SO}^{2}}$ and eigenenergies $E_{a/b}=\mp\Delta/2$.
Note that these doublets follow the general structure of
time-reversed pairs in Eq.~\eqref{eq:kramerspair}. In the presence
of a magnetic field, we shall neglect the quadratic term in
\eqref{eq:HB} altogether. This amounts to assuming the dot to be
much smaller than the magnetic length, i.e.
$eB\ell_{dot}^{2}\ll\hbar$, which will then ensure that $e
B^{2}\langle r^{2}\rangle-e\langle({\bf r}\cdot{\bf
B})^{2}\rangle\ll\langle({\bf r}\times{\bf p})\cdot{\bf B}\rangle$.
The remaining linear terms in Eq.~(\ref{eq:HB}) have the following
matrix elements in the original four-state basis:
\begin{equation} \label{eq:HZHL}
\langle i\sigma|H_{B}|j\sigma'\rangle=
\frac{g_{0}\mu_{B}}{2}\tau^{0}_{ij}\mathbf{B}\cdot
\boldsymbol{\tau}_{\sigma\sigma'}+i\mu_{B}\tau^{y}_{ij}
\tau^{0}_{\sigma\sigma'}\boldsymbol{\lambda}\cdot\mathbf{B},
\end{equation}
where $\boldsymbol{\lambda}=-i\langle 1|\mathbf{L}|2\rangle$ is a
(real) vector characteristic for the two levels. Expressing this in
the zero-field eigenbasis ($a,b$ doublets), we can now find the low
field splitting of the individual doublets. That is, for fields low
enough that this splitting will be much smaller than $\Delta$, we
obtain the effective single-doublet ($m=a,b$) Hamiltonians:
\begin{equation}
\langle m\eta|H_{B}|m\eta'\rangle=\frac{g_{0}\mu_{B}}{2}BZ_{m}
(\sin\xi_{m},0,\cos\xi_{m})\cdot\boldsymbol{\tau}_{\eta\eta'}
\end{equation}
where we have defined
\begin{equation}
Z_{a/b}=\sqrt{\frac{\delta^{2}\sin^{2}\zeta}{\delta^{2}+\Delta_{SO}^{2}}%
+\left(\cos\zeta\pm\frac{2\tilde{\lambda}\Delta_{SO}}{g_{0}\Delta}%
\right)^{2}},\label{eq:Z}
\end{equation}
with $\sin\xi_{a/b}=\pm\delta(\sin\zeta)/(Z_{a/b}\Delta)$ and where
$\tilde{\lambda}=|\boldsymbol{\lambda }|(\cos\zeta\cos\theta+\sin
\zeta\cos\varphi\sin\theta)$ denotes the projection of
$\boldsymbol{\lambda}$ on $\mathbf{B}$ in terms of the relative
angles $(\zeta,\theta,\varphi$) between the two intrinsic vectors,
$\mathbf{B}_{SO}$ and $\boldsymbol{\lambda}$, and the external
$\mathbf{B}$, all indicated in Fig.~\ref{fig:CondPlots}(a). The
eigenstates of this hamiltonian are readily found by a rotation
within the plane spanned by $\mathbf{B}_{SO}$ and $\mathbf{B}$, i.e.
$|\widetilde{m\eta}\rangle^{(0)}=\mathbb{R}_{y}(\xi_{m})|m\eta\rangle$,
where
$\mathbb{R}_{y}(\xi)=\tau^{0}\cos(\xi/2)+i\tau^{2}\sin(\xi/2)$.
Apart from this splitting, a finite field will also mix the $a$ and
$b$ doublets. Starting from this last eigenbasis, which only refers
to the direction of $\mathbf{B}$, we include this mixing to linear
order in the magnitude $|\mathbf{B}|$ from first order perturbation
theory, i.e. $|\widetilde{a\eta}\rangle^{(1)}=
|\widetilde{a\eta}\rangle^{(0)}+\langle
b\eta'|H_{B}|a\eta''\rangle[\mathbb{R}_{y}(\xi_{a})]_{\eta\eta''}
|b\eta'\rangle^{(0)}$, where $\langle b\eta'|H_{B}|a\eta\rangle=
(\tilde{\lambda}\delta\tau^{z}_{\eta'\eta}-
g_{0}\sin(\zeta)\Delta_{SO}\tau^{x}_{\eta'\eta}/2)\mu_{B}B/\Delta$.
The amplitudes for tunneling into this lowest lying Zeeman-split
Kramers doublet $|\widetilde{a\eta}\rangle^{(1)}$, are now readily
found using (\ref{eq:ab0}). Notice that it is only this last
non-unitary mixing of $a$ and $b$ doublets which prevents us from
diagonalizing $t_{\alpha a}^{\eta\eta'}$ by a canonical
transformation of the conduction electrons.
\noindent


Whereas the SOI could not be discerned at zero field, the
angle-dependent bias voltage asymmetry of $dI/dV$, confirmed by our
simple model in Figs.~\ref{fig:CondPlots}(b-f), is a unique
signature of SOI at finite field. Such bias asymmetries have often
been found in experiments on various quantum
dots.~\cite{Jespersen06,Schmauss09} Nevertheless, it is often
difficult to rule out the influence of incipient charge
fluctuations, setting in at slightly higher bias voltages, as the
source of this asymmetry. The only unambiguous evidence for such
SOI-induced bias asymmetry will therefore be the observation of its
variation with a change in the direction of the magnetic field.
Taken together with the possible angle dependence of the Zeeman
splitting itself (cf. e.g. Refs.~\onlinecite{Takahashi09,Katsaros10}),
such a measurement can thus reveal otherwise inaccessible details on
the spin-orbit coupling in a given quantum dot.


We thank V. Golovach and P. Brouwer for stimulating discussions and
acknowledge financial support from the Danish Agency for Science,
Technology and Innovation and from the European Union under the FP7
STREP program SINGLE (J.P., K.F.).


\end{document}